\documentclass[a4paper,12pt]{article}
\usepackage{a4wide}
\usepackage{verbatim}
\usepackage{bm,latexsym,amsmath,amssymb,amsfonts,mathrsfs,amsthm,mathtools}
\usepackage{bbm} 
\usepackage{graphicx}
\usepackage[nosort]{cite}
\usepackage{color}
\usepackage{latexsym}
\usepackage{epsf}
\usepackage{slashed} 
\usepackage{array} 
\usepackage{booktabs}
\usepackage{subcaption}

\renewcommand{\d}{\textrm{d}}

\newcommand{\ba}{\begin{equation}\begin{aligned}}
\newcommand{\ea}{\end{aligned}\end{equation}}

\renewcommand{\d}{\textrm{d}}

\def\calo         {{\cal O}}

\def\reals        {{\mathbb R}}
\def\integers          {{\mathbb Z}}

\newcommand{\del}{\partial}

\DeclareMathOperator{\sech}{sech}

\DeclareMathOperator{\arcsinh}{arcsinh}

\allowdisplaybreaks

\allowdisplaybreaks[3]

\usepackage[hang]{footmisc}
\usepackage[compress,square,numbers]{natbib}
\usepackage[colorlinks,linktocpage,linkcolor=blue,citecolor=blue,urlcolor=blue]{hyperref}

\renewcommand{\d}{\mathrm{d}}

\begin{document}

\numberwithin{equation}{section}
\setcounter{tocdepth}{2}

\begin{flushright}
CTPU-PTC-22-07
\end{flushright}

\vfil
\vspace*{-.2cm}
\noindent

\vspace*{1.2cm}

\begin{center}
	
	{\LARGE \bf{Axion wormholes with massive dilaton} }
	
	\vspace{2 cm} {\large Stefano Andriolo$^a$, Gary Shiu$^b$, Pablo Soler$^c$,   Thomas Van Riet$^d$}\\
\vspace{1 cm} {\small\slshape $a$ Department of Physics, Ben-Gurion University of the Negev,\\ Beer-Sheva 84105, Israel}\\
{\small\slshape $b$ Department of Physics, University of Wisconsin-Madison,\\ Madison, WI 53706, USA}\\
{\small\slshape $c$  Center for Theoretical Physics of the Universe,\\ Institute for Basic Science, Daejeon 34051, South Korea}\\
{\small\slshape $d$ Instituut voor Theoretische Fysica, K.U.Leuven, \\Celestijnenlaan 200D, B-3001 Leuven, Belgium}\\
\vspace{.8cm} {\upshape\ttfamily stefanoa@post.bgu.ac.il, shiu@physics.wisc.edu, soler@ibs.re.kr thomas.vanriet@kuleuven.be }\\

\vspace{2cm}

{\bf Abstract}

\begin{quotation}
If Euclidean wormholes contribute meaningfully to the path integral of quantum gravity they can have important implications for particle physics and cosmology. The dominant effects arise from wormholes whose sizes are comparable to the cut-off scale of effective field theory, for which ultraviolet corrections become relevant. We study corrections to classical axion wormhole solutions in string motivated scenarios in which the dilaton partner of the axion becomes massive. We find corrections near the neck region which are consistent with a recent version of the weak gravity conjecture for axions.
\end{quotation}
\end{center}
\newpage

\section{Introduction}
Axions coupled to gravity allow for non-trivial Euclidean saddle points with finite action that correspond to wormhole geometries~\cite{Giddings:1987cg}. Such instantons are argued to break the classical shift symmetry of the axion and hence violate axion charge conservation by means of baby universes, carrying charge, that leave or enter the mother universe \cite{Coleman:1988cy}. They can hence lead to very interesting phenomenology, as reviewed in~\cite{Hebecker:2018ofv}. In the context of the Swampland program \cite{Brennan:2017rbf, Palti:2019pca, vanBeest:2021lhn}, in particular regarding global symmetries \cite{McNamara:2020uza} and the Weak Gravity Conjecture (WGC) \cite{ArkaniHamed:2006dz}, axionic wormholes have been the subject of many recent works.  As wormholes couple electrically to axion scalars they have been considered as the natural objects to study for extensions of the WGC to axions and instantons~\cite{Montero:2015ofa, Brown:2015iha, Heidenreich:2015nta, Hebecker:2016dsw, Hebecker:2017wsu,Hebecker2017} (see~\cite{Harlow:2022gzl} for a recent review and references).

At the same time, contributions from general wormhole geometries to the effective action give rise to several apparent paradoxes which can be made particularly sharp in the context of holography~\cite{Maldacena:2004rf, ArkaniHamed:2007js, Katmadas:2018ksp}. Whether wormhole instantons should really contribute to the path integral of quantum gravity is an old question with differing viewpoints~\cite{Maldacena:2004rf, ArkaniHamed:2007js, VanRiet:2020pcn, McNamara:2020uza, Eberhardt:2021jvj,Saad:2021uzi,Heckman:2021vzx,Blommaert:2021fob}. In this more general context, axionic wormholes seem prime examples because of their explicitness and simplicity. For instance in a holographic context, they trigger vevs of marginal operators which makes holographic identifications more straightforward.\footnote{Other kinds of Euclidean wormholes have been studied in a holographic context in \cite{Betzios:2019rds, Betzios:2021fnm,Marolf:2021kjc,Maldacena:2004rf}} A simple way to resolve the puzzles could arise if wormholes were perturbative unstable, in the sense that quadratic fluctuations around the wormhole saddle could have negative modes that lower the Euclidean action \cite{Hertog:2018kbz}. These could indicate that wormholes should not contribute significantly to the path integral. However, recent work \cite{Loges:2022nuw} has shown that, with the appropriate boundary conditions for the gauge invariant perturbations that keep the axion charge of the wormhole fixed, there are no such negative modes and thus the axion wormhole in \cite{Giddings:1987cg} is perturbatively stable.

In string theory compactifications axions tend to pair up with dilatons. When enough supersymmetry is preserved both the axion and the dilaton remain massless at tree-level. There exist then three classes of Euclidean solutions depending on the relative size of the axion and dilaton momenta~\cite{Gutperle:2002km}. When the axion momentum dominates, one finds again the wormhole geometry. When the axion-momentum exactly balances the dilaton momentum, the geometry is flat (in the absence of a cosmological constant). When the dilaton momentum dominates, the solution is singular and its physical significance is still unclear. There are various reasons to think of these 3 classes as ``over-extremal", ``extremal" and ``sub-extremal" as reviewed in \cite{VanRiet:2020csu}. For instance, the extremal solution has a concrete incarnation as the supersymmetric D-instanton in string theory \cite{Gibbons:1995vg} and has a charge equal to its on-shell action such that there is a ``Euclidean no-force condition" \cite{ VanRiet:2020pcn}. Further arguments using the dimensional reduction and the c-map \cite{Bergshoeff:2004fq, Brown:2015iha, Heidenreich:2015nta}, probe actions \cite{VanRiet:2020pcn} and charge/action ratios suggest this picture.\footnote{The wormhole, however, is really a dipole solution since it describes two different spaces glued together, with the axion charge having opposite signs when measured on opposite sides of the wormhole. In that sense it is not entirely correct to call this ``over-extremal" since it is not smoothly connected to the extremal solution which describes only 1 universe.} 

In the absence of dilatons there only exist wormholes and this notion of extremality ceases to exist. In string theory set-ups, dilatons can be absent in an effective field theory sense; their masses $m$ are above the cut-off scale $\Lambda_{UV}$.  In particular, string  compactifications relevant for phenomenology require moduli stabilisation, after which one may expect classically massless axions and massive dilatons. Whether three classes of instantons still exist in these scenarios is unclear and this question is one of the motivations behind the present work. Somewhat surprisingly,  not much work seems to have been done in this direction beyond the original studies of~\cite{Kallosh:1995hi} (see however~\cite{Alvey:2020nyh,Hamaguchi:2021mmt}). In contrast, the most top-down constructions of axion wormholes \cite{Giddings:1989bq, Bergshoeff:2004pg, Hertog:2017owm} feature massless dilatons, making them phenomenologically useful only indirectly.

The most dominant wormholes are those with the smallest action and so one is lead to contemplate small wormholes, with neck radii close to $\Lambda_{UV}^{-1}$.  It is therefore crucial to understand the effect of corrections to the EFT on them.
The effect of higher derivative interactions on relatively small wormholes were studied in~\cite{Andriolo:2020lul} and inspired the authors to formulate a specific WGC: corrections to the wormhole backgrounds are conjectured to decrease the action-to-charge ratio. This was explicitly verified for the dominant higher derivative terms under certain assumptions. Integrating out massive fields with $m>\Lambda_{UV}$ typically generates higher derivative corrections in the EFT, but one can also simply study the EFT at a higher energy scale and include the massive fields explicitly. This is one of the goals of this work. But, aside from verifying the WGC conjecture of \cite{Andriolo:2020lul}, we would like to better understand the space of instanton solutions in theories with massive dilatons.

Not surprisingly massive dilatons correct the wormhole geometry around the neck at distance scales below $m^{-1}$ and this can be computed rather explicitly. We also find that the conjecture of \cite{Andriolo:2020lul} is nicely obeyed. Since moduli are stabilized in realistic string compactifications, the corrections from massive dilatons studied in this work can be 
the most relevant ones in a phenomenological context.

This paper is organised as follows. We describe the axion-dilaton-gravity system in Section \ref{setup}, and review the three different types of Euclidean solutions (cored solutions, flat solutions, and wormholes) when the dilaton is massless. We then turn to the massive dilaton case in Section \ref{massive_dilaton} which naturally led us to consider two opposite regimes of dilaton mass. Analytic results can be obtained when the dilaton mass is large compared with the inverse wormhole size. We present the perturbatively corrected wormhole solution and its Euclidean action in Section \ref{sec:analytics}. When the dilaton mass is small compared with the inverse wormhole size, we have to resort to numerical methods. We present our numerical results for the wormhole solution and its Euclidean action in Section \ref{sec:numerics}. We found that the action-to-charge ratio of wormholes increases monotonically with their charge (i.e. with their size), giving further support to the axion WGC. We conclude in Section \ref{conclusion}.

\section{The setup}\label{setup}
\setcounter{equation}{0}

Consider the axio-dilaton-gravity (ADG) system with a non-trivial potential for the dilaton. After Wick-rotating to Euclidean signature and carefully dualizing the axion into a 2-form $B$ with field strength $F=dB$, the action reads  \cite{Giddings:1987cg,Giddings:1989bq,GHS,Rey:1989xj}\footnote{See also reviews \cite{Hebecker:2018ofv,Hebecker:2016dsw}, and in particular \cite{Collinucci:thesis} for a thorough explanation of the dualization procedure.}
\begin{align}
\label{Euclidean_Ead}
S_E=\int \d^4 x \sqrt{g}
\bigg[
- \frac{M_P^2}{2} R + \frac{1}{2}(\del_\mu\phi)^2 + \frac{1}{12 f^2} e^{-\frac{\beta}{M_P}\phi} F^2
+ V(\phi)
\bigg]\,.
\end{align}
The exponential coupling between the dilaton and the 3-form flux (with $\beta \geq0$) arises generically in string compactifications, where the dilaton $\phi$ corresponds typically to the string coupling or to a geometric modulus. Moduli stabilization is an involved process which leads in general to complicated dilaton potentials and which may interact non-trivially with the axion sector. We will assume in this work that~\eqref{Euclidean_Ead} accurately describes a consistent EFT below an energy cutoff scale $\Lambda_{UV}\lesssim M_P$, with a dilaton potential of the simple form
\begin{align}
V(\phi) = m^2 \phi^2 \,,
\end{align}
where $m\ll \Lambda_{UV}$. Notice that this potential explicitly breaks the $SL(2,\reals)$ symmetry of the $m=0$ case. We will not address the question of how such an EFT may arise from a UV complete setting, in particular from an explicit string compactification. Henceforth, we will set $M_P=1$, explicitly recovering it where needed.

\subsection{Equations of motion and Ansatz}
The equations of motion of the system~\eqref{Euclidean_Ead} are given by:
\begin{align}
\label{EOMs}
\nonumber 
0&=\del_\mu\left(\sqrt{g}e^{-\beta\phi}F^{\mu\nu\rho}\right)
\,,\\
\nonumber 
0&=\frac{\beta}{12 f^2}e^{-\beta\phi} F^2
+\frac{1}{\sqrt{g}}\del_\mu\left(\sqrt{g}g^{\mu\nu}\del_\nu\phi\right)
- \del_\phi V
\,,\\
0&=R_{\mu\nu}
-  \del_\mu\phi\del_\nu\phi
-  V g_{\mu\nu}
+ \frac{1}{6 f^2} e^{-\beta\phi} g_{\mu\nu} F^2
- \frac{1}{2 f^2} e^{-\beta\phi} F_{\mu\rho\sigma}F_\nu{}^{\rho\sigma}
\,.
\end{align}
To these, one has to add the Bianchi Identity $\d F=0$.

\medskip

We are interested in this work in spherically symmetric solutions of the form
\begin{align}
\label{GS_Ansatz}
\d s^2 = h(r)^2 \d r^2 + a(r)^2 \d\Omega_3^2\,, 
\quad\,\, 
F = q \varepsilon \,,
\quad\,\, 
\phi(r) & \,,
\end{align}
where $d\Omega_3^2$ and $\epsilon$ are the line element and volume form of the unit three-sphere, respectively. At fixed $r=r_0$, the metric reads $ds^2_{r_0}=a(r_0)^2 d\Omega_3^2$ and so $a(r_0)$ represents the physical radius of the solution (not to be confused with the radial coordinate $r_0$). Without loss of generality we can focus on $a(r)>0$.
We look for solutions that are asymptotically flat and have vanishing dilaton, namely
\begin{align}
\d s^2 \to \d r^2 + r^2 \d\Omega_3^2 \,, 
\quad 
\phi\to0 \,, 
\qquad \text{with} ~~ r \to \infty \,.
\end{align}

The axion charge (corresponding to the $F$-flux) contained in a three-sphere is given by the quantized parameter $q$ as
\begin{align}
\int_{S^3} F=2\pi^2 q \in\integers\,.
\end{align}
We will take for concreteness $q>0$, the most general results can be obtained by replacing $q\to|q|$.

\medskip
With the above Ansatz the equation of motion and Bianchi identity for $F$ are trivially satisfied. The dilaton equation is 
\begin{align}
\label{eqdilaton}
0 = D(a^3 D\phi) 
- 2 a^3 m^2 \phi
+ \frac{\beta q^2}{2 f^2} \frac{e^{-\beta  \phi }}{a^3} 
\,,
\end{align}
and the Einstein equations are
\begin{align}
\label{ee1}
0&=
- m^2 \phi^2
+ \frac{q^2}{ f^2} \frac{e^{-\beta\phi }}{a^6}
- 3 \frac{D^2a}{a}
- (D\phi)^2 \,,\\
\label{ee2}
0 &= 
\frac{1}{a} D(a^2Da)
- 2
+ a^2 m^2 \phi^2
\,,
\end{align}
where $Df(r)\equiv \tfrac{f'}{h}$, with primes denoting derivatives with respect to the radial coordinate $r$. We will make use throughout this work of different radial coordinates (different 	`gauge choices') which allow us to fix either $a$ or $h$ (or a combination thereof). Different choices will be convenient for different computations and interpretations of results.

\subsection{Solutions with massless dilaton}\label{solutions}

\medskip
For $m=0$ solutions to the ADG system are very well known \cite{Gutperle:2002km, Bergshoeff:2004fq, Hebecker:2018ofv, Hebecker:2016dsw}. In the gauge $a(r)=r$, they are:
\begin{itemize}
\item \emph{Cored} solutions:
\begin{align}
\label{cored}
h = \left( 1 + \frac{c^4}{r^4} \right)^{-1/2} \,,
\qquad
e^{\beta\phi} = \frac{q^2}{6 f^2 c^4} \sinh^2\left[ K + \frac{\sqrt{3}}{2\sqrt{2}}\, \beta \arcsinh \left( \frac{c^2}{r^2} \right) \right] 
\,,
\end{align}
with
\begin{align}\label{constc}
c^4 = \frac{q^2}{6 f^2} \sinh^2 K
\,,
\end{align} 
where $K$ is an arbitrary integration constant determining a family of solutions. The physical significance of $K$ can be understood by embedding the system into a higher dimensional theory. Cored solutions arise from higher dimensional Euclidean branes that wrap internal cycles of the geometry. $K$ is then related to the tension and charge of such objects. The simplest case would be that of an {\it under-extremal} black hole wrapping the compactification circle of a 5d theory, where $K$ is determined by the black hole charge and mass~\cite{Hebecker:2016dsw,Heidenreich:2015nta}.

Both the metric and the dilaton profiles are singular at $r=0$ in the cored solution and hence cannot be fully described within EFT. The solution~\eqref{cored} should therefore only be trusted in the regime in which the metric curvature is smaller than the UV cutoff of the theory ($r\gg 1/\Lambda_{UV}$) and the dilaton takes values well described in the EFT.

\item \emph{Flat} solutions:
\begin{align}
\label{flat}
h = 1 \,,
\qquad
e^{\beta\phi} =  \left( 1 + \frac{q \beta}{4 f r^2} \right)^2 
\,,
\end{align}
These can be understood as the limit $K\to 0$ of the cored solutions. From a higher dimensional perspective, they correspond to an {\it extremal} brane wrapping a compactification cycle, e.g. a 5d extremal Reissner-Nordstrom black hole whose Euclidean worldline wraps the internal circle.  

The metric is flat in these solutions, but the dilaton blows up as $r\to 0$, which is generally expected to signal a breakdown of the EFT. While one can give a full description of these objects in certain UV settings (e.g. when they represent BPS D-brane instantons), one should be cautious when treating the small $r$ regime in the ADG effective theory. 

\item \emph{Wormhole} solutions:  These exist only for couplings $\beta$ smaller than a critical value, $0\leq \beta<\beta_c\equiv\frac{2\sqrt{2}}{\sqrt{3}}$~\cite{Giddings:1987cg} and have
\begin{align}
\label{semi_wh}
h = \left( 1 - \frac{a_0^4}{r^4} \right)^{-1/2} \,,
\qquad
e^{\beta\phi} = \frac{q^2}{6 f^2 a_0^4} \cos^2\left[\frac{\beta}{\beta_c} \arccos \left( \frac{a_0^2}{r^2} \right) \right] 
\,,
\end{align}
where $a_0$ is given by
\begin{align}
a_0^4 = \frac{q^2}{6 f^2} \cos^2\left(\frac{\pi}{2}\frac{\beta}{\beta_c} \right)
\,.
\end{align}
The metric has a coordinate singularity at $r=a_0$ but the geometry is smooth, as can be seen by changing coordinates to $r^2\equiv a_0^2 \cosh(2\tilde{r})$ so that
\begin{align}\label{fullcoordinate}
ds^2=a_0^2 \cosh(2\tilde{r})\left(d\tilde{r}^2 + d\Omega^2_3\right)\,.
\end{align}
This is manifestly $\mathbb{Z}_2$-symmetric under $\tilde{r}\to - \tilde{r}$ and extends smoothly throughout the range $-\infty < \tilde{r} < \infty$. The geometry is that of a wormhole with two asymptotically flat regions $\tilde{r} \to \pm \infty$ connected by a spherical throat that takes its minimal radius $a_0$ at $\tilde{r}=0$. 
The action of the wormhole can be easily computed and yields
\begin{align}
\label{ADG_action_value}
S_{m=0} = \frac{8 \pi^2 q M_P}{\beta f} \sin \left( \frac{\pi}{2}\frac{\beta}{\beta_c} \right) \,.
\end{align} 
This is the action of the entire wormhole $-\infty<\tilde{r}<+\infty$.

Wormholes can be well described within EFT as long as the curvature remains well below the UV cutoff scale. This is guaranteed when the charge $q$ is  sufficiently large so that the neck radius $a_0$ is macroscopic, i.e. such that $a_0 \gg \frac{1}{\Lambda_{UV}}$. On the other hand, as already mentioned, wormholes solutions exist in the limited range of couplings $0\leq \beta<\beta_c$, while cored and flat solutions arise for any $\beta>0$. It is interesting that the latter solutions have a divergent action in the limit $\beta \to 0$ in which the dilaton decouples (see e.g.~\cite{Bergshoeff:2004fq}), or equivalently when the dilaton becomes infinitely massive. Hence, in this limit, the only macroscopic solutions that contribute to the effective action are the wormhole ones.
 
It is also worth noticing that the dilaton typically traverses significant distances along the wormhole, $\Delta \phi= \phi(r=a_0)-\phi(\infty)=\phi(r=a_0)$, as can be seen directly from~\eqref{semi_wh}: 
\begin{align}\label{newdist}
e^{\beta\Delta\phi}=\cos^{-2}\bigg(\frac{\pi}{2}\frac{\beta}{\beta_c}\bigg)
\end{align}
Interestingly, the distance is independent of the wormhole charge (i.e. the wormhole size), and only depends on $\beta$. We present this dependence in Figure~\ref{fig:massless_displacement}. As $\beta\to \beta_c$ the dilaton profile diverges and the wormhole solution ceases to exist. For $\beta \gtrsim 0.92$ the distance becomes trans-Planckian, raising concerns on the validity of the solution within the EFT. The distance decreases as $\beta$ gets smaller, and becomes zero at $\beta=0$, where the dilaton decouples. We will see in Section~\ref{sec:numerics} how these results are affected when the dilaton gains a mass.

\begin{figure}[htb]
\begin{center}
\includegraphics[scale=0.5]{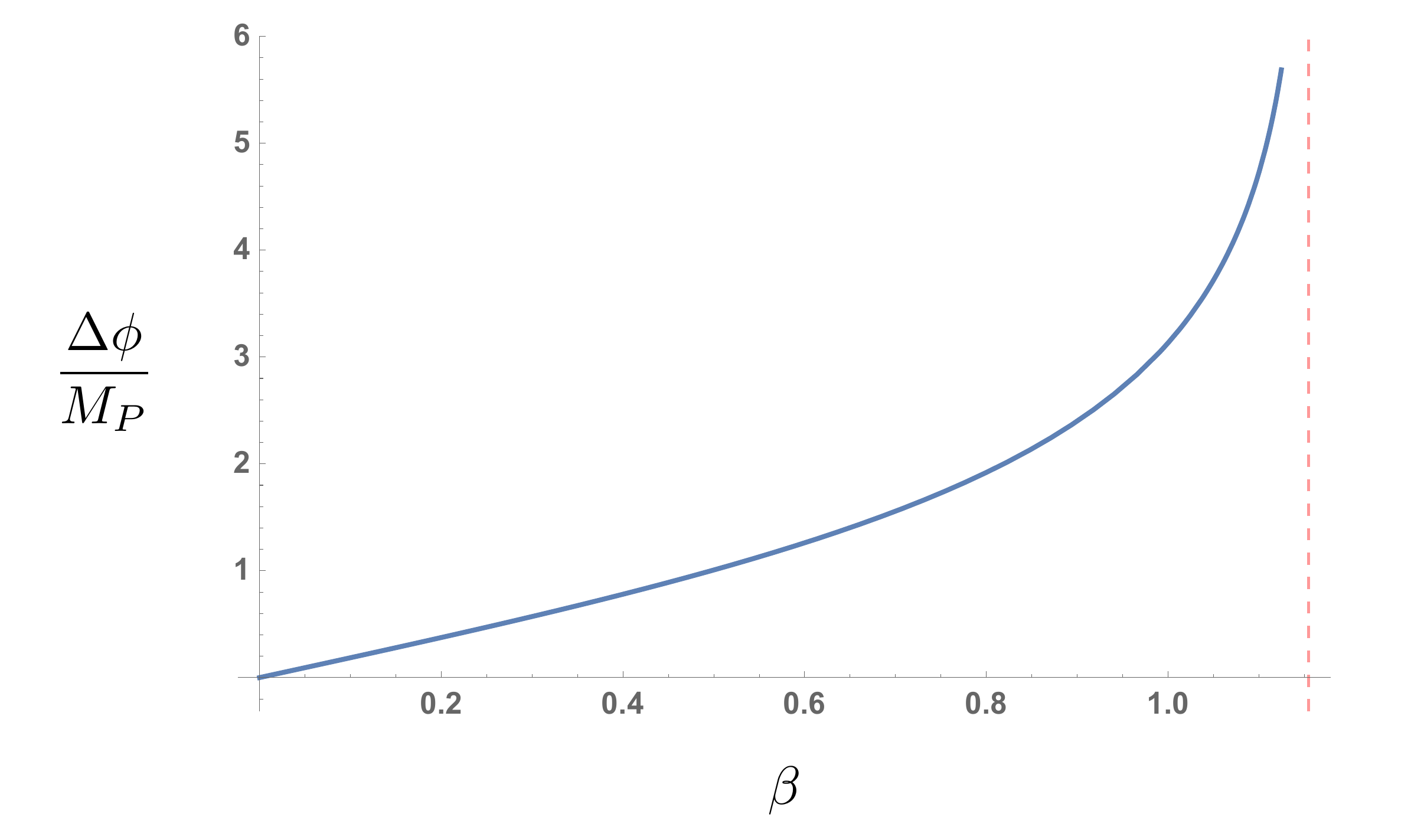}
\caption{\label{fig:massless_displacement}
\textit{\small{Distance traversed by the dilaton, in Planck units, along the wormhole $\Delta \phi= \phi(r=a_0)-\phi(\infty)=\phi(r=a_0)$ as a function of the axion-dilaton coupling $\beta$. Dashed lines represent $\beta_c$ (red) and the transplanckian value $\beta \approx 0.92$ (blue) respectively.}}}
\end{center}
\end{figure}

\end{itemize}

\section{The massive dilaton system}\label{massive_dilaton}

The previous statements motivate us to focus in the following on the behavior of wormhole solutions when the dilaton gets stabilized. Before presenting the analysis, let us make a few comments about cored and flat solutions.

In the case $m\neq0$, the equations of motion immediately tell us that exactly flat solutions ($h=1,\, a=r$) are not viable any more. One can in fact refine this statement and argue that there are no solutions with geometries that are smooth at the origin $r\to 0$ (in the $r$-coordinate, with $a(r)=r$). For that, notice first that a smooth curvature $R=\frac{6}{r^2} - \frac{6}{r^2 h^2} + \frac{6 h'}{r h^3}$ requires $h=1 + \calo(r^2)$ as $r\to0$. As a consequence, the $rr$ component of the Einstein tensor, $G_{rr}=\frac{3}{r^2} - \frac{3 h^2}{r^2}$, is also smooth at the origin. This means that both the trace $T$ and the $T_{rr}$ (times $2/h^2$) component of the stress-energy tensor 
\begin{align} 
T &= 
\frac{q^2 e^{-\beta\phi}}{2\, f^2\, r^6} - 2 m^2 \phi^2 - \frac{1}{2} \frac{\phi'^2}{h^2}
\,, \\
\frac{2}{h^2} T_{rr} &= 
-\frac{q^2 e^{-\beta\phi}}{2 \,f^2 \, r^6} -  m^2 \phi^2 + \frac{1}{2} \frac{\phi'^2}{h^2} \,,
\end{align}
must be smooth as well. However, these quantities differ only by $\sim m^2\phi^2$, thus one concludes that $\phi$ must be smooth at the origin. Therefore, $\phi'$ must cancel the singularity produced by $r^{-6}$ in $T$ and $T_{rr}$. The latter is only true if $\phi$ is logarithmic, yielding a contradiction.

\begin{figure}[htb]
\begin{center}
\hspace{-23pt}
\includegraphics[width=0.54\textwidth]{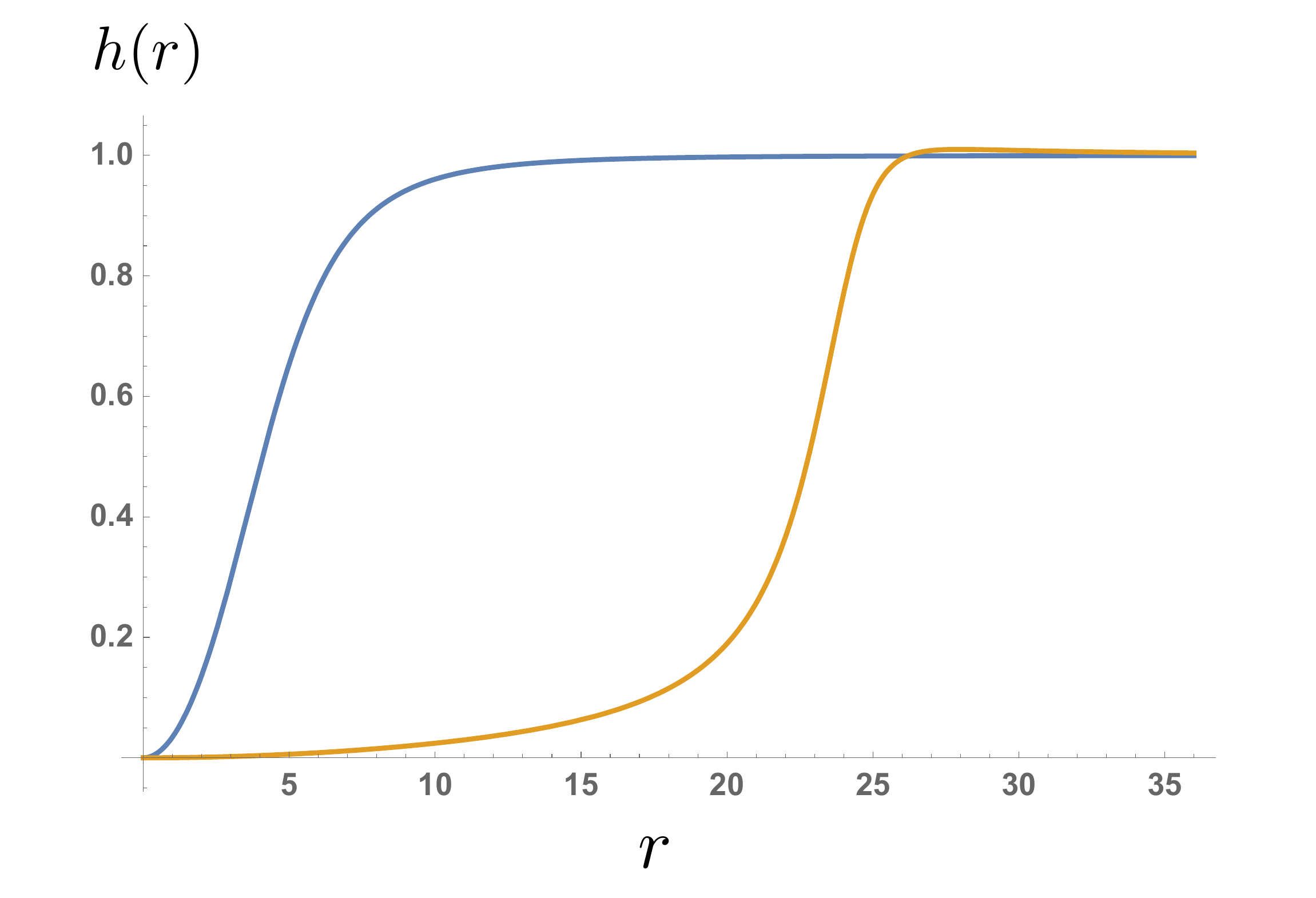}
\hspace{-25pt}
\includegraphics[width=0.54\textwidth]{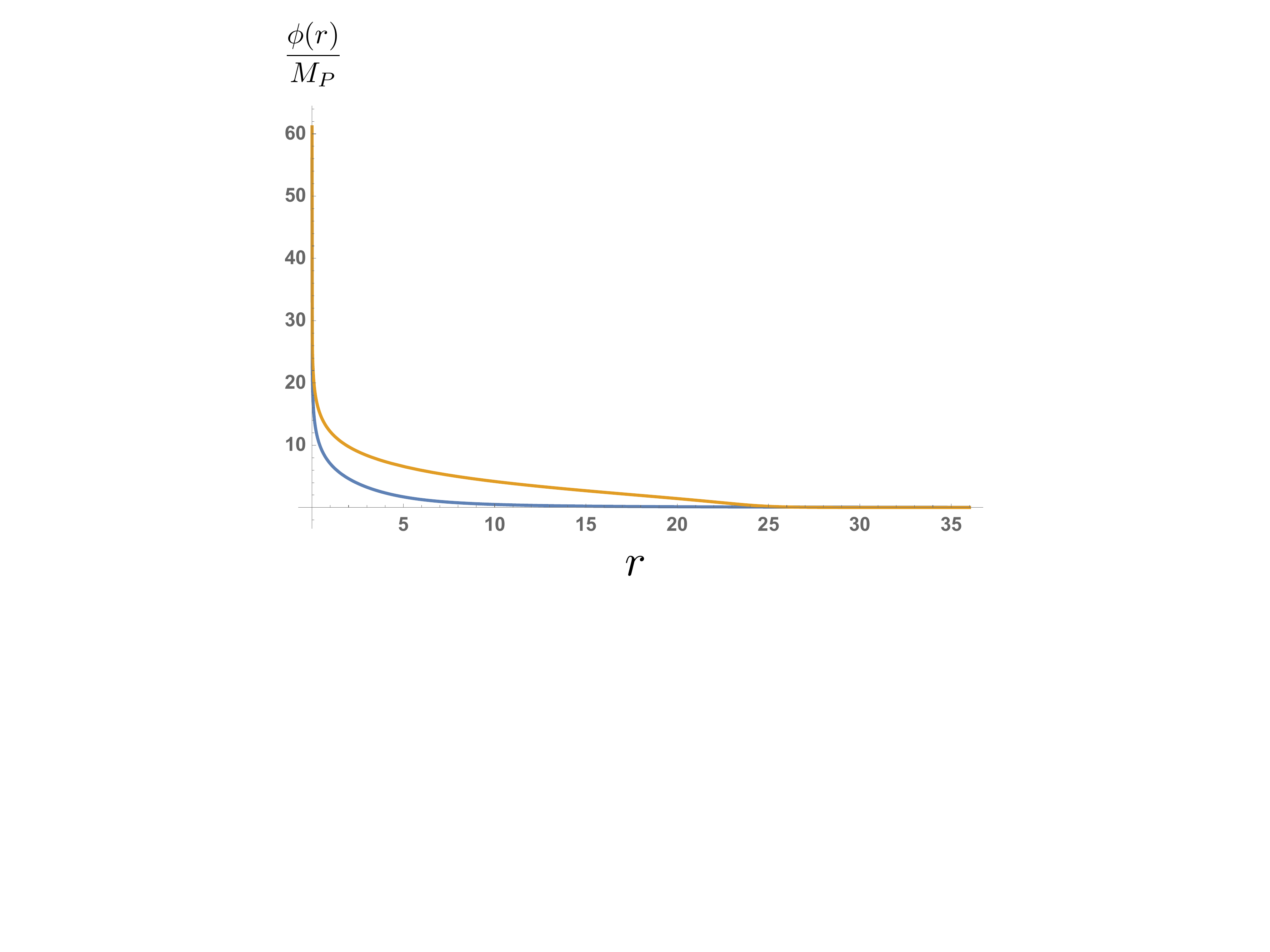}
\caption{\label{fig:coredsols}
\textit{\small{
Examples of cored instanton solutions for $m=0, K=1$ (blue) and $m=0.2$ (green), $m=0.5$ (purple), $m=0.75$ (red). Here, $\beta=1$ and $\tfrac{q}{f}=10$. We obtained the massive solution by integrating the equations starting from the asympotics (here at $r\approx 15$) where we imposed asymptotically flat boundary conditions.}}}
\end{center}
\end{figure}

Deformations of cored solutions induced by the dilaton mass can be computed numerically. An example is presented in Fig.~\ref{fig:coredsols}. The singular behavior of the solution near the origin, however, prevents us from computing the action and so we move in the rest of this note to study wormhole solutions, which are well behaved in the EFT.

\medskip
We are interested in wormhole solutions of the system~\eqref{Euclidean_Ead} with finite $m$. The massless limit $m\to 0$ should reproduce the results of previous sections. The opposite extreme, $m\to \infty$, in which the dilaton freezes is also well understood. Analytic solutions are easily derived, and can in fact be obtained from the $m=0$ solutions by taking the limit $\beta \to 0$ in which the dilaton and axion decouple. In particular, the action for wormholes in the axion-gravity system (with no dilaton) is simply $S_{m\to\infty} =\frac{\sqrt{6}\,\pi^3 q M_P}{f}$.

What we really mean by the limit $m\to 0$ (respectively $m\to \infty$) is of course that the dilaton mass is much smaller (larger) than the relevant energy scale of the solution, which is given by the inverse neck radius, i.e. $m^2 \ll a_0^{-2} \sim f/q$ (respectively $m^2 \gg f/q$). We can hence equivalently refer to the zero charge $q\to 0$ (respectively $q\to \infty$) limit, which is more appropriate when comparing wormholes of different sizes within a given EFT with fixed dilaton mass. 

The key quantity to study when addressing the WGC for wormholes and axions is the action-to-charge ratio
\begin{align}
s_q\equiv \frac{f\, S_q}{q \,M_P}\,.
\end{align}
One may expect, and we will indeed confirm, that the action-to-charge ratio of wormholes of finite charge $q$ interpolates between the two limiting cases $q\to 0$ and $q\to \infty$:
\begin{align}\label{hierarchy}
 \frac{8 \pi^2}{\beta} \sin \left( \frac{\pi}{2}\frac{\beta}{\beta_c} \right)=s_0 < s_q  < s_\infty=\sqrt{6}\, \pi^3
\,.
\end{align}
The upper bound in~\eqref{hierarchy} was introduced in~\cite{Hebecker:2016dsw,Hebecker:2018ofv} as a version of the WGC applied to axions and wormholes.\footnote{The action-to-charge ratio was computed in~\cite{Hebecker:2016dsw,Hebecker:2018ofv} for a semi-wormhole in terms of the quantized charge~$n=2\pi^2 q$, which gives $\tfrac{S_{n}f}{2 n M_P}\to\tfrac{\pi\sqrt{6}}{4}$.} It tells us that, in a given theory with fixed dilaton mass, smaller wormholes have a smaller action-to-charge ratio than larger ones. 

This confirms in a specific setup the general findings of~\cite{Andriolo:2020lul}, which derived the behavior~\eqref{hierarchy} induced by the leading higher derivative corrections to the effective action. Indeed, we could regard the reduction of the action-to-charge ratio we observe as coming from higher derivative corrections that arise from integrating out the massive dilaton. The leading contributions studied in~\cite{Andriolo:2020lul} are sufficient to address the action-to-charge ratio reduction for large wormholes, such that $a_0^{-1} \ll m$. Our results go beyond and explore the regime of smaller wormholes with $a_0^{-1} \gtrsim m$.

Nevertheless, we should stress that our analysis is only valid for wormholes that are sufficiently large to be well described in EFT, i.e. such that $a_0^{-1}\ll \Lambda_{UV}$. In this sense, our results still support a {\it mild} version of the WGC but cannot confirm or refute stronger versions that deal with truly microscopic wormholes. 

We should also caution the reader from interpreting the upper bound in~\eqref{hierarchy} as an {\it extremality} bound. The notion of extremality for wormholes and gravitational instantons is in general poorly understood. Setups with enough supersymmetry or where a relation to higher dimensional extremal objects is available (e.g. in the case of BPS D-brane instantons) suggest that the flat solutions presented in the previous section are to be interpreted as extremal. Wormholes could be then regarded as {\it super-extremal}.\footnote{This interpretation is supported by the fact that the action of probe D-instantons in a wormhole background is minimized when they sit far away from the wormhole neck (i.e. D-instantons are ``repelled'' by wormholes)~\cite{VanRiet:2020pcn}.} Nevertheless, the fate of flat solutions is unclear once the dilaton gains a mass and the notion of extremality becomes obscure. Regardless of these subtleties, we can regard~\eqref{hierarchy} as the WGC-like statement that smaller wormholes have a smaller action-to-charge ratio than larger ones.

We proceed to study in detail the behavior of wormhole solutions and their action-to-charge ratios when the dilaton becomes massive. The regimes of very large wormholes $a_0^{-1}\ll m$ and small wormholes $a_0^{-1}\gtrsim m$ can be treated separately. The former is amenable to a perturbative analysis while the latter requires numerical methods.

\section{Large wormholes, $ma_0\gg1$}
\label{sec:analytics}

Very large wormholes (i.e. wormholes with a very large charge $q$) whose curvature is everywhere much smaller than the dilaton mass probe only energy scales much smaller than $m$. One can, to leading order, integrate out the dilaton and obtain an analytic solution in which $\phi$ is frozen:
\begin{align}
h\approx \left(1-\frac{q^2}{6 f^2 r^4}\right)^{-1/2}\,,\qquad \phi\approx 0\, .
\end{align}
Their action-to-charge ratio is given by $s_\infty=\sqrt{6}\, \pi^3$ and their neck radius $a^2_0\approx \frac{q}{\sqrt{6}f}$ is much larger than the inverse dilaton mass.

We will perform a perturbative expansion in the small parameter $\epsilon =\frac{f}{qm^2}\ll 1$ which controls the ratio of the inverse dilaton mass to the neck radius of large wormholes. The solution can then be written as
\begin{align}
h(r)^{-2}=1-\frac{q^2}{6f^2 r^4}\left[1+\epsilon\, \delta h(r) +{\cal O}(\epsilon^2) \right]\,, \qquad \phi(r)= \epsilon \,\delta \phi(r) +{\cal O}(\epsilon^2)
\end{align}
The equations of motion~\eqref{eqdilaton}-\eqref{ee2} can be easily solved at order $\epsilon$ and yield
\begin{align}
\delta h(r)=-\frac{q^3 \beta^2}{8 f^3 r^6}\, , \qquad \delta \phi(r)= \frac{q^3 \beta}{4 f^3 r^6}\,.
\end{align}
Notice that these corrections indeed satisfy $\delta h(r), \delta\phi(r)\lesssim {\cal O}(1)$ since the coordinate system used (i.e. the gauge $a(r)=r$) covers only the region $r^2 \gtrsim \frac{q}{\sqrt{6}f}$. The corrected solutions can then be written as
\begin{align}\label{perts}
h(r)^{-2}=1-\frac{q^2}{6f^2 r^4}\left[1- \frac{q^2 \beta^2}{8m^2 f^2 r^6} +{\cal O}\left(\frac{f^2}{q^2 m^4}\right) \right] \, \qquad \phi(r)=\frac{q^2\beta}{4m^2 f^2 r^6}+{\cal O}\left(\frac{f^2}{q^2 m^4}\right) \,. 
\end{align}

This perturbed solution can be brought to the more appropriate coordinate system~\eqref{fullcoordinate} which covers the full wormhole. In this gauge, the metric takes the form $ds^2=r(\tilde{r})^2(d\tilde{r}^2+d\Omega_3^2)$\,. The coordinate transformation hence satisfies $rd\tilde{r}=h(r)dr$ which can be solved to order ${\cal O}(\epsilon)$ as
\begin{align}
r(\tilde{r}) = \left(\frac{q^2}{6 f^2}\right)^{1/4}  \cosh^{1/2}(2 \tilde r)
\left[1 + \epsilon \, \delta r(\tilde r) +{\cal O}(\epsilon^2) \right]  \,,
\end{align}
where the function $\delta  r(\tilde{r})$ can be written as
\begin{align}
\delta r(\tilde{r}) &= \frac{-3\sqrt{3}}{8\sqrt{2}} \,\beta^2 \sech(2\tilde{r}) ~{}_2F_1\left[-\frac{1}{2},2,\frac{1}{2},-\sinh^2(2\tilde{r})\right] \nonumber\\
&= \frac{3\sqrt{3}}{16\sqrt{2}} \,\beta^2 \sech(2\tilde{r})
\left[ - 3  + \sech^2(2\tilde{r}) - 6 \sinh(2\tilde{r})\arctan\left(\tanh(\tilde{r})\right)  \right] \,.
\end{align}
The dilaton is obtained from~\eqref{perts} simply as
\begin{align}
\phi(\tilde{r})=\epsilon \frac{q^3\beta}{4f^3 r(\tilde{r})^6}+{\cal O}(\epsilon^2)=\frac{3\sqrt{6} f \beta}{2\,m^2 q}\sech^3(2\tilde{r})+{\cal O}\left(\frac{f^2}{q^2 m^4}\right) \,.
\end{align}

The metric $ds^2=r(\tilde{r})^2(d\tilde{r}^2+d\Omega_3^2)$ is smooth throughout the whole wormhole range $-\infty <\tilde{r} < \infty$. The $\mathbb{Z}_2$ symmetry $\tilde{r}\to -\tilde{r}$ is manifest, which allows us to easily identify the point $\tilde{r}=0$ as the location of the wormhole neck, which has a radius
\begin{align}
a_0=r(0)=\left(\frac{q^2}{6f^2}\right)^{1/4}\left(1-\frac{3\sqrt{6}}{16}\frac{f \beta^2}{qm^2}+{\cal O}(\epsilon^2)\right)\,.
\end{align}
Notice that the corrections lower the radius of the wormhole neck.  

The quantity we are most interested in is the wormhole action-to-charge ratio. Upon using the equations of motion, the action can be written as
\begin{align}
\nonumber\label{action}
S_q &=  \int \d^4x \sqrt{g} 
\left[ \frac{1}{6 f^2} e^{-\beta\phi} F^2 - m^2 \phi^2  \right] \\
&=
2 \pi^2 \int \d r h a^3 
\left[ 
\frac{q^2}{f^2} \frac{e^{-\beta\phi}}{a^6} - m^2 \phi^2 
\right]
\,, 
\end{align}
where in the last step we used the generic Ansatz \eqref{GS_Ansatz} and integrated over $S^3$ (notice that there are no contributions from boundary terms). The integral can be performed in the $\tilde{r}$-coordinates (i.e. in the gauge $a=h$) which covers the full wormhole, and to order ${\cal O}(\epsilon)$ yields the action-to-charge ratio
\begin{align}\label{pertresult}
s_q=\frac{S_q f}{q M_P} 
&=
 \sqrt{6} \pi^3 
\left( 
1 - \frac{\sqrt{6}}{2\,\pi}  \frac{\beta^2 f M_P}{q \,m^2}  
\right)+{\cal O}\left(\frac{f^2\,M_P^2}{q^2 \,m^4}\right)
\,.
\end{align}
where we have reinstated $M_P$ explicitly. 

As mentioned before, the negative correction to the action-to-charge ratio in~\eqref{pertresult} can be interpreted as a (mild) version of the WGC. The action-to-charge ratio is larger for wormholes with bigger charge, and asymptotes to $\sqrt{6}\,\pi^3$ in the infinite charge limit ($\epsilon\to 0$). Of course, our results so far are only valid for wormholes with curvature smaller than the scale $m^2$, and hence can be viewed as a confirmation of the general analysis of~\cite{Andriolo:2020lul} in the concrete setup~\eqref{Euclidean_Ead}. We would like next to extend our results to the regime of smaller wormholes in which the $\epsilon$-expansion is no longer appropriate.

\section{Small wormholes: $ma_0\lesssim \calo(1)$}
\label{sec:numerics}

The previous analysis is not applicable to wormholes whose curvature at the neck is comparable to or bigger than the dilaton mass. This occurs for small charges $\frac{q\,m^2}{f} \lesssim {{\cal O}(1)}$. In this case we can divide the full wormhole geometry into two regions, one region near the wormhole neck where curvature is significant, and another asymptotic region where spacetime is almost flat and the dilaton mass is dominant (of course there are two equivalent asymptotic regions, one at each side of the wormhole). In the following we find numerical solutions to the equations of motion near the wormhole neck and match them to analytic solutions valid asymptotically.

Our strategy begins by integrating numerically the equations of motion starting from the neck (c.f.~\cite{Kallosh:1995hi}). We focus on equations~\eqref{eqdilaton} and~\eqref{ee2}. We find it useful to work in yet another radial coordinate $\tau$ in which $h(\tau)=1$. This gauge choice makes the $\mathbb{Z}_2$-symmetry $\tau\to -\tau$ manifest and covers the full wormhole $-\infty <\tau < \infty$, with the neck located at $\tau=0$. Two boundary conditions are fixed by requiring smoothness of the solution at the neck:
\begin{align}
\label{smoothness}
\phi'(0) = a'(0) = 0 \,,
\end{align}
where primes denote derivatives with respect to $\tau$. The remaining boundary conditions $\phi(0)\equiv \phi_0$ and $a(0)\equiv a_0$ are not independent, they are related through the equations of motion~\eqref{ee1} and~\eqref{ee2} as
\begin{align}
m^2\phi_0^2+\frac{q^2}{2f^2}\frac{e^{-\beta \phi_0}}{a_0^6}-\frac{3}{a_0^2}=0\,.
\end{align}
We can further restrict our choices of boundary conditions by demanding that the wormhole solution has a familiar shape, namely
\begin{align}
\phi''(0)<0 \,,  
\qquad 
a''(0)>0 \,,
\end{align}
which implies that
\begin{align}
m \, a_0\,  \phi_0  < \sqrt{2} \,, 
\qquad
a_0^2\,m^2\left(\phi_0^2+\frac{2\phi_0}{\beta}\right) < 3  \,,
\end{align}
as can be derived from the equations of motion~\eqref{eqdilaton}-\eqref{ee2}.

We study numerical solutions to the equations of motion as a function of the boundary conditions subject to the above constraints. We look for those boundary values that yield solutions which approach flat spacetime at large distances, at which the numerical solution can be glued to the asymptotic analytic one. At large $\tau$ the metric approaches $a(\tau)\to \tau$, and the dilaton has a profile of the form
\begin{align}
\phi(\tau) \stackrel{\tau\to\infty}{\longrightarrow} \frac{\beta q^2}{4m^2 f^2 \tau^6} 
+ C_1  \frac{e^{\sqrt{2}m\tau}}{\tau^{3/2}}
+ C_2  \frac{e^{-\sqrt{2}m\tau}}{\tau^{3/2}}
\,.
\end{align}
The exponentially decaying third term is of course irrelevant at large distances and should be ignored. The exponentially growing second term, on the other hand, is unacceptable, and so requires a choice of boundary conditions ($\phi_0$) such that $C_1=0$ and the dilaton decays at large distances.

\subsection{The action-to-charge ratio and dilaton displacement}

Using this `shooting method' we can find a numerical wormhole solution for any choice of  $q$, $f$, $m$ and $\beta$. Much of the dependence on these parameters is however redundant. As before, our main interest will be in the action-to-charge ratio $s_q$ and the maximum dilaton displacement $\Delta\phi=\phi(0)-\phi(\infty)=\phi_0$. It turns out that these quantities depends on $q$, $f$ and $m$ only through the combination $\tfrac{qm^2}{f}$. This can be seen, e.g. in the gauge $a(\tilde{r})=h(\tilde{r})$, as follows. The equations of motion~\eqref{eqdilaton}-\eqref{ee2} are invariant under the rescaling $q\to k\,q$, $m\to \tfrac{m}{\sqrt{k}}$ and $a\to\sqrt{k}a$ for any parameter $k$. Under this transformation the action scales as $S\to kS$, as can be seen directly from~\eqref{action}. Hence, the action-to-charge ratio of wormholes related by this rescaling satisfy
\begin{align}
s_q(m)=\frac{S_q(m) f}{q\, M_P}=\frac{S_{kq}(\tfrac{m}{\sqrt{k}}) f}{kq\, M_P}=s_{kq}(\tfrac{m}{\sqrt{k}})=s_{m^2q}(1)=s_1(m\sqrt{q})\,.
\end{align}
Of course, the dependence on the axion decay constant (the inverse gauge coupling) $f$ is only through the ratio $q/f$, and so we find the desired result that $s_q$ depends only on the combination~$\tfrac{qm^2}{f}$. This can be seen explicitly in the perturbative expansion~\eqref{pertresult}. A similar conclusion follows immediately for the dilaton displacement, since $\phi$ is invariant under the $k$-transformation.

We can hence without loss of generality fix any value of the dilaton mass and probe different values of the wormhole charge $\tfrac{q}{f}$. We can just as well, and this is what we do in practice, fix $\frac{q}{f}=\sqrt{2}$ and solve the equations for different values of $m$. It is furthermore convenient and illustrative to express $m$ in units of the inverse neck radius $a_0^{-1}$, the product $ma_0$ tells us whether wormholes are large or small with respect to the dilaton (inverse) mass scale (notice that $ma_0$ is invariant under the above $k$-rescaling, a property that holds also on the gauge $h(\tau)=1$).

Finally, we will set the remaining parameter $\beta=\sqrt{2}$ for concreteness when doing explicit evaluations. Recall that not all values of $\beta$ allow for wormhole solutions (c.f. Section~\ref{solutions}). Our numerical analysis suggests that the critical maximum value $\beta_c$ that allows for wormholes monotonically increases with $m$, starting from $\beta_{c}(m=0)=\tfrac{2\sqrt{2}}{\sqrt{3}}\simeq 1.63 $. This behaviour naturally aligns with the fact that for $m\to\infty$ any value of $\beta\geq 0$ admits wormhole solutions. Thus, we can take any $\beta<\tfrac{2\sqrt{2}}{\sqrt{3}}$, for which wormholes exist at any $m$.

\begin{figure}[!h]
\begin{center}
\includegraphics[scale=0.45]{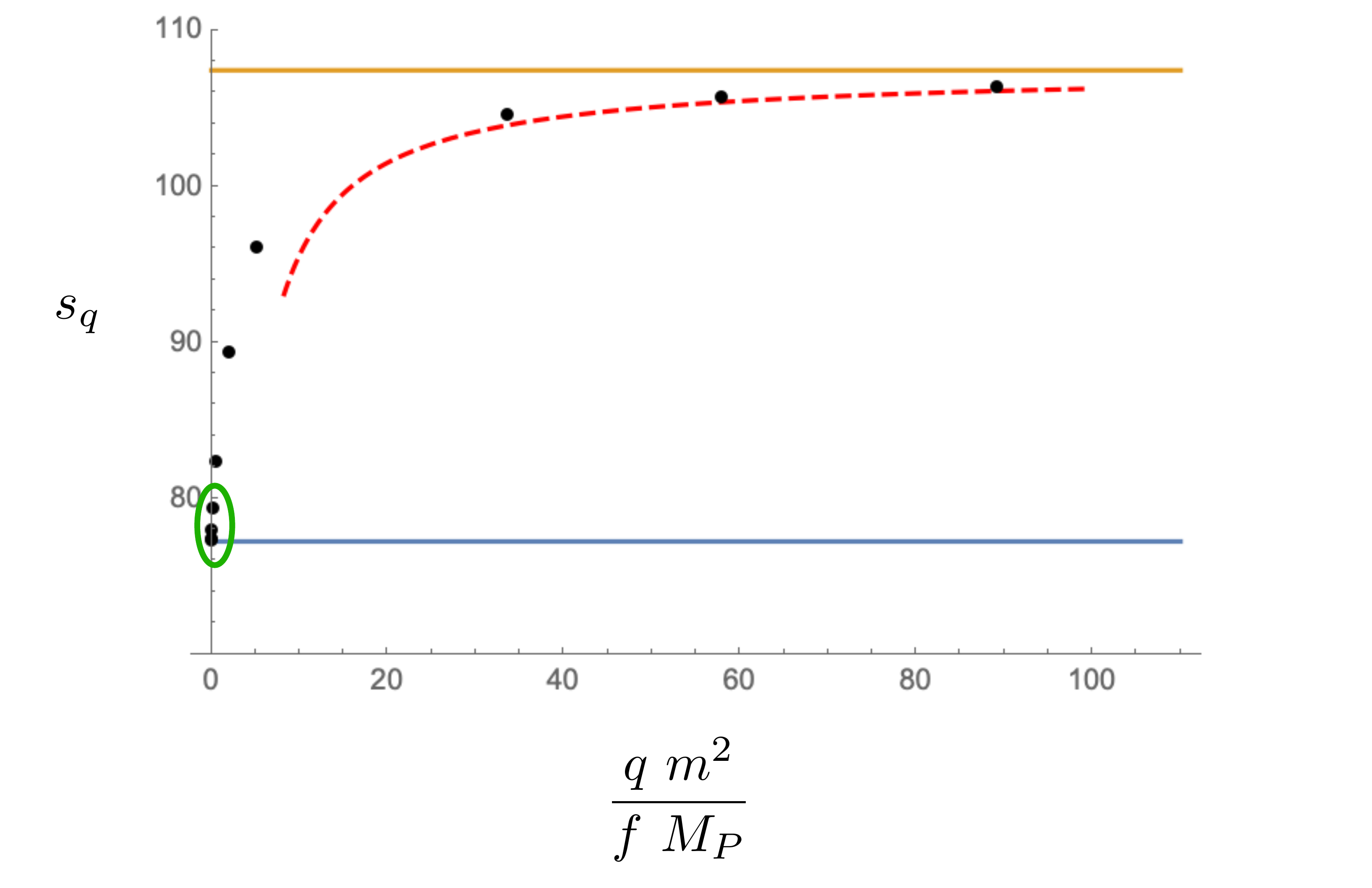}
\caption{\label{fig:results}
\textit{\small{Behaviour of the charge-to-mass ratio $s_q=\tfrac{S_q f}{q\,M_P}$ as a function of $\frac{q \,m^2}{f\, M_P}$ (for $\beta=\sqrt{2}$). Small wormholes have action values close to the one of the massless dilaton system (blue line), while the action of bigger wormholes tends to the one of the system without a dilaton (orange line). The dashed red line represents the analytical result~\eqref{pertresult} for $a_0 m \gg 1 $ obtained in the previous Section. We have verified that also the points cummulating in the green encircled area behave monotonically. }}}
\end{center}
\end{figure}

\begin{figure}[!h]
\begin{center}
\includegraphics[scale=0.45]{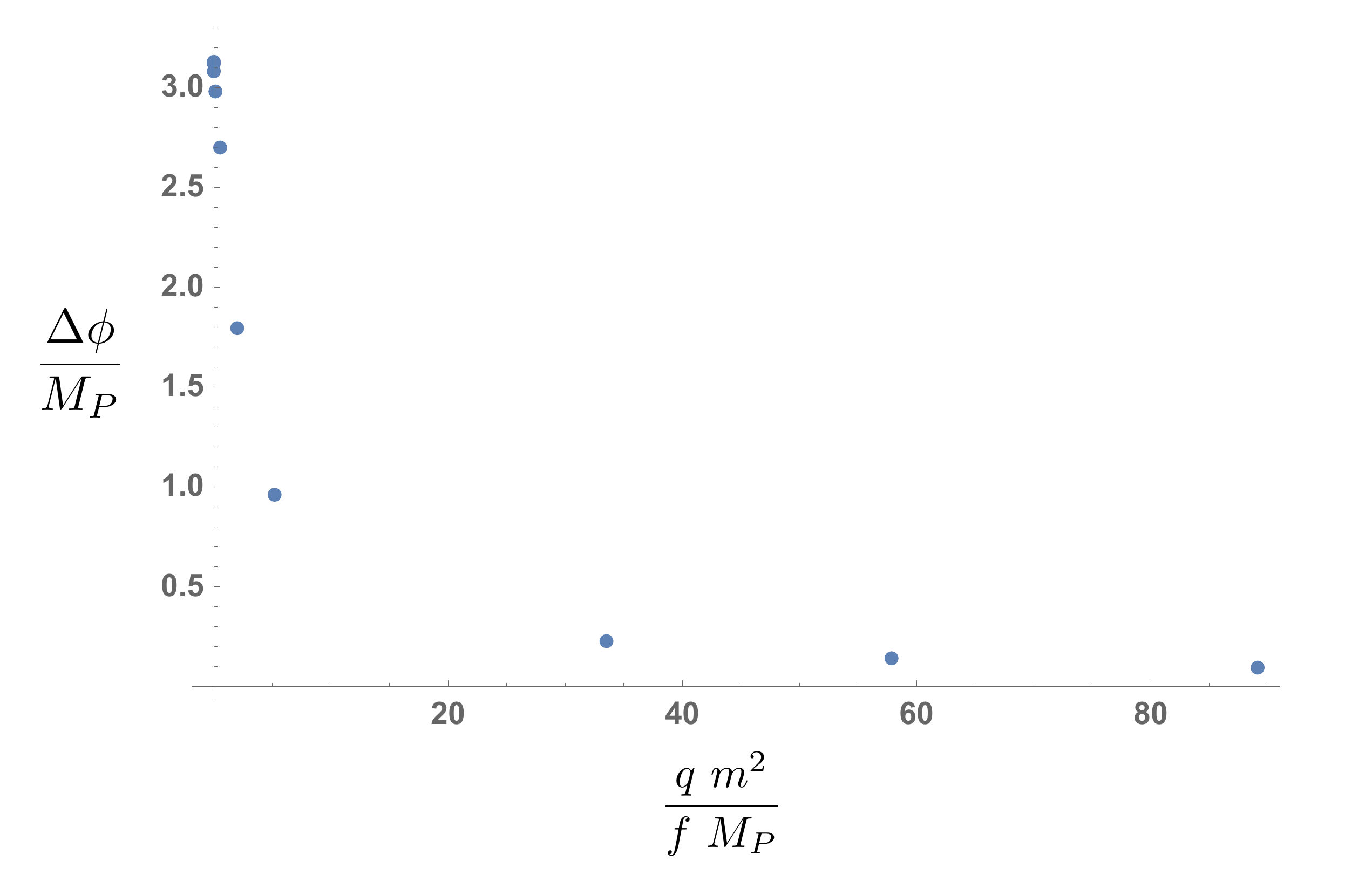}
\caption{\label{fig:massive_distance}
\textit{\small{Distance traversed by the dilaton along the wormhole, $\Delta \phi= \phi_0$, as a function of the wormhole size, parametrized by $\tfrac{qm^2}{fM_P}$ (for~$\beta=\sqrt{2}$). }}}\end{center}
\end{figure}

Our results are summarized in Table \ref{tab:results} and in Figures~\ref{fig:results} and~\ref{fig:massive_distance}, where we can see how the action-to-charge ratio is larger for bigger wormholes. They interpolate between the analytically solvable cases of a massless dilaton (small wormholes, $s_0=4\sqrt{2}\,\pi^2\sin(\tfrac{\sqrt{3}\pi}{4})=54.599$) to the case of infinite mass in which the dilaton decouples (large wormholes, $s_\infty=\sqrt{6}\,\pi^3=75.950$). As described above, we can interpret this as a version of the WGC for wormholes.

\begin{table}
\centering
\resizebox{\columnwidth}{!}{%
\begin{tabular}{c|ccccccccccc|c}
\toprule
$ma_0$ & 0 ($m=0$) & $\frac{1}{100}$ & $\frac{1}{50}$ & $\frac{1}{20}$ & $\frac{1}{10}$ & $\frac{1}{5}$ & $\frac{1}{2}$ & 1 & 3 & 4 & 5 & $\infty$ ($\phi=0$)  \\
\midrule
$s_q$ & 54.599& 54.638& 54.727& 55.156& 56.129& 58.260& 63.157& \
67.922& 73.949& 74.720& 75.213& 75.950 \\
$m^2q/f$  & 0  & 0.0012& 0.0047& 0.0285& 0.1079& 0.3686& 1.403& 3.644& 23.72& 40.89& \
63.02  & $\infty$  \\
$\phi_0$ & 2.215& 2.212& 2.207& 2.178& 2.105& 1.907& 1.271& 0.680& 0.1599& \
0.1015& 0.0698 & 0 \\
\bottomrule
\end{tabular}
}
\caption{\label{tab:results} \textit{\small{Results of the numerical evaluation of wormhole solutions (for $\beta=\sqrt{2}$). The first line is the neck radius in terms of the dilaton mass, $a_0m$. As discussed above, this is our input parameter. The following lines are the  action-to-charge ratio $s_q=\tfrac{S_q f}{q}$, the parameter $\tfrac{m^2 q}{f}$, and the boundary dilaton value $\phi_0$. }}}
\end{table}

Turning on a mass induces a dependence of the dilaton displacement $\Delta \phi=\phi_0$ on the wormhole charge, which is shown in Fig.~\ref{fig:massive_distance}. As expected, the distance is small for large wormholes and approaches zero in the infinite charge (i.e. infinite dilaton mass) limit in which the dilaton freezes. For smaller wormholes, the displacement grows and becomes transplanckian at $\tfrac{qm^2}{fM_P}\lesssim 2.1$ (obtained via extrapolation), which corresponds to $ma_0\lesssim 1$ (c.f. Table~\ref{tab:results}). In the limit $q\to 0$ the result approaches that  of the massless dilaton system, $\Delta \phi\approx 2.2 M_P$, which matches the results of Section~\ref{solutions} (for $\beta=\sqrt{2}$). As mentioned there, the large displacement of the dilaton may raise concerns over the wormhole solutions, but we see from Figure~\ref{fig:massive_distance} that the non-zero mass improves the behavior of the dilaton profile on large wormholes.

\section{Conclusion}\label{conclusion}
We have studied the corrections to the Euclidean axion wormholes of Giddings and Strominger \cite{Giddings:1987cg} caused by a massive dilaton partner. When the mass is high (i.e. for large wormholes) this corrects the pure axion wormhole and when the mass is small (i.e. for small wormholes) this corrects the wormhole solution supported by massless axions and dilatons~\cite{Giddings:1989bq, Gutperle:2002km, Bergshoeff:2004fq}. These setups are generic and particularly relevant for phenomenological applications of string theory, in which dilatons are typically stabilised at tree level, while axions can remain perturbatively massless. It will be interesting to explore such phenomenological implications in more detail, e.g. in the context of inflation, (ultra-light) dark matter, the strong CP-problem, or for axions in the micro-eV mass range (see e.g.~\cite{Hebecker:2018ofv} for a review of such potential applications). We have deferred such discussions to future work and focussed here on more formal conceptual aspects of Euclidean wormholes.

We have performed a perturbative analysis for large $ma_0$ and a numerical analysis for small $ma_0$, with $a_0$ the wormhole neck radius. The two results together point to a consistent picture, whose main results are summarized in Table \ref{tab:results} and Figure \ref{fig:results}. In particular we have found that, for fixed $m,f,\beta$:
\begin{align}
\label{WGCGS}
& \lim_{q\to\infty} \left(\frac{S_q f}{q M_P} \right)
= \sqrt{6} \pi^3 \,, \\
\label{BHs_analogy}
&\frac{\d }{\d q} \left(\frac{S_q f}{q M_P} \right) > 0 
\,.
\end{align}
These equations reflect the WGC expectation that smaller wormholes have a smaller action-to-charge ratio. There is a close analogy with the way black hole charge-to-mass ratios change in the presence of corrections \cite{Kats:2006xp,Hamada:2018dde,Cheung:2018cwt,Bellazzini:2019xts,Aalsma:2019ryi,Loges:2019jzs,Loges:2020trf,Cano:2019oma,Cano:2019ycn,Jones:2019nev,Cremonini:2019wdk,Aalsma:2020duv,Cremonini:2020smy,Arkani-Hamed:2021ajd,Aalsma:2021qga}, and agrees with previous findings dictated by positivity conditions \cite{Andriolo:2020lul}. It implies that $S(q_1+q_2)>S(q_1)+S(q_2)$, meaning that a wormhole of charge $q_1+q_2$ contributes less to the path integral than a pair of wormholes with charges $q_1$ and $q_2$. This fact may be interpreted as a ``wormhole instability'', analogous to the extremal black hole instability motivating the standard version of the WGC.

We have furthermore studied the behavior of the dilaton displacement $\Delta\phi$ along wormhole geometries. In the massless case, the displacement is independent of the size of the wormhole and changes only with the axion-dilaton coupling $\beta$. For $\beta\to 0$, the dilaton decouples and $\Delta \phi\to 0$, while for $\beta \gtrsim 0.92$ the displacement becomes transplanckian (and blows up as $\beta \to \beta_{crit}$), raising concerns on the validity of the solutions. The behaviour changes significantly in the presence of a mass, which induces a dependence of $\Delta \phi$ on the wormhole size, as shown in Figure~\ref{fig:massive_distance}. The dilaton displacement decreases with the size of the wormhole and tends to zero asymptotically.

Our results should help clarify the issue of wormhole stability. Note that pure axion wormholes were found to be perturbatively stable \cite{Loges:2022nuw}, despite earlier claims \cite{Hertog:2018kbz}. Still the pertubative stability of wormholes supported by massless axion and dilatons remains unclear, and that situation is directly relevant for the holographic embedding of axion wormholes \cite{Hertog:2017owm} with all their associated paradoxes \cite{Maldacena:2004rf, ArkaniHamed:2007js, Katmadas:2018ksp}. On the other hand, setups with massive dilatons are the most relevant for phenomenology, and the stability analysis of \cite{Loges:2022nuw,Andriolo:2020lul} together with the results of this paper indicate that the wormholes only feature ``non-perturbative" instabilities in the sense that wormhole fragmentation will dominate the path integral (see also \cite{VanRiet:2020pcn}). What this means for the truly dominating saddle points remains unclear and finding the answer to this question might require an understanding of the full UV completion, not just the leading corrections to the EFT.

\section*{Acknowledgments}We would like to thank Takanori Anegawa, Arthur Hebecker, Norihiro Iizuka, Tommy Krug, Aitor Landete, Gregory Loges, Toshifumi Noumi, Min-Seok Seo, Toshiaki Takeuchi and Brecht Wagemans for useful discussions. The work of GS is supported in part by the DOE grant DE-SC0017647. The work of PS was supported by the Institute for Basic Science under the project code, IBS-R018-D1. The work of TVR is supported by the KU Leuven C1 grant ZKD1118C16/16/005. The work of SA is supported by the Israel Science Foundation (grant No. 741/20) and by the German Research Foundation through a German-Israeli Project Cooperation (DIP) grant "Holography and the Swampland".

\bibliography{WGC}
\bibliographystyle{utphys}

\end{document}